# Towards Responsible and Fair Data Science:
# Resource Allocation for Inclusive and Sustainable Analytics

*Genoveva Vargas-Solar*
*Centre National de la Recherche Scientifique CNRS, LIRIS UMR 5205*
*Campus la Doua, Villeurbanne 69622, France*
genoveva.vargas-solar@cnrs.fr
http://www.vargas-solar.com

## Abstract

*This project addresses the challenges of responsible and fair resource allocation in data science (DS), focusing on DS queries[1] evaluation. Current DS practices often overlook the broader socio-economic, environmental, and ethical implications, including data sovereignty, fairness, and inclusivity. By integrating a decolonial perspective, the project aims to establish innovative fairness metrics that respect cultural and contextual diversity, optimise computational and energy efficiency, and ensure equitable participation of underrepresented communities. The research includes developing algorithms to align resource allocation with fairness constraints, incorporating ethical and sustainability considerations, and fostering interdisciplinary collaborations to bridge gaps between technical advancements and societal impact. This work aims to reshape into an equitable, transparent, and community-empowering practice challenging the technological power developed by the Big Tech[2]. The project aligns to the new trends willing to redefine the production model and scientific context of the database community and to the scientific project of the LIRIS lab.*

## Keywords

*Fairness-aware data science, responsible AI and analytics, data science queries, multi-objective optimization, decolonial methodologies, sustainable resource allocation*

## Context and motivation

The rapid development of artificial intelligence (AI)-based data analytics methods has generated significant momentum but also raised concerns about the immense computing, economic, and ecological resources required for training models and deploying final solutions. The resource demands are colossal, leading to substantial ecological and economic impacts. Additionally, the bias and incompleteness inherent in data and (numerical and AI) models trained on partial content—often in English, digital, and predominantly from the Global North—exacerbate epistemic violence [3] by marginalizing diverse perspectives and knowledge systems from the Global South[4]. The evolution of technology, which promotes the automation of knowledge production under the guise of being ethically agnostic, is reshaping the distribution of economic, political, and cultural power. This shift represents a new form of colonialism driven by technological dominance and concentrated monopolies. Decolonisation efforts now extend to include Europe and other regions in the Global North that lack technological independence.

---

[1] From a database perspective, my previous research introduced the concept of a data science query, hypothesising that DS pipelines can be treated as queries. In this framework, tasks function as operators applied to data, evaluated step by step, with the execution order governed by a control flow that may include loops.

[2] *"Big Tech, also known as the Tech Giants or Tech Titans, is a grouping of the largest IT companies in the world. It typically refers to the Big Five United States tech companies: Alphabet, Amazon, Apple, Meta, and Microsoft, or the Magnificent Seven, which includes Nvidia and Tesla. Big Tech can also include Chinese companies such as Baidu, Alibaba, Tencent, and Xiaomi (BATX)"*, https://en.wikipedia.org/wiki/Big_Tech (accessed the 9th of January 2025).

[3] Epistemic violence refers to the harm done to individuals or groups through the marginalisation, silencing, or devaluation of their knowledge, experiences, and ways of understanding the world. It occurs when certain forms of knowledge, often those from dominant or powerful groups, are privileged over others, leading to the suppression or erasure of alternative perspectives. This can happen in various contexts, such as academia, media, and technology, where the voices and knowledge of marginalised communities are ignored, misrepresented, or dismissed.

[4] From postcolonial and decolonial perspectives, the global North is often associated with whiteness and characterised by Western-, Euro-, and American-centric ideologies, while the global South is perceived as the "other," positioned as divergent from the hegemonic "norm."[23]

The use of underlying infrastructure, such as computing servers, storage locations, and governance, is often obscured from users. Critical questions arise regarding the origins and handling of data: Where does the data originate? When the data involves specific territories, communities, or individuals, are they aware of its use for analytics purposes, and do they consent to it? Furthermore, do they agree with the interpretations and conclusions drawn from the analytics? Are they informed about where the hardware processing their data is located, and do they consent to that as well? Finally, are they aware of, and do they approve of, the vast amount of computing resources consumed by AI models to analyse their data?

In relation to the above questions, this project explores the challenges, existing solutions, and open issues related to resource allocation in data science (DS) queries enactment environments that implement scientific data-driven experiments into *in-silico* processes. The focus is on how to allocate resources effectively while adhering to service level objectives (SLOs[5]) and fairness requirements, which include factors such as server location, data provenance, energy consumption, sovereignty, carbon footprint, and economic cost. The goal is to optimise resource distribution across different stages of the DS queries evaluation process within a given architecture, ensuring that these fairness criteria are integrated into the allocation strategy. Additionally, the objective is to curate (i.e., document) the decision-making strategies used for selecting and allocating computing resources, identifying their providers, and tracing the data and individuals involved in their collection. This ensures responsibility and accountability throughout the processes. This approach aligns with decolonial methodologies, aiming to promote more sustainable, equitable, and responsible alternatives to the resource-intensive AI practices commonly employed in contemporary DS.

> Over the next five years, my research project will focus on the evaluation of DS queries, building on my previous work and integrating enhanced considerations of responsibility, fairness, and inclusion. Positioned at the forefront of emerging perspectives in the database discipline, this project adopts a transdisciplinary, DEI-aware, and responsibility-driven approach. It aligns with and complements the scientific objectives of the database group at LIRIS[6].

### Towards inclusive and fair data science: vision and problem statement

Inclusion in the context of DS entails integrating human involvement throughout the responsible, ethical, and accountable design and evaluation processes of DS queries on target platforms. Advanced DS techniques provide quantitative tools for analytical and predictive tasks, leveraging mathematical models, statistics, and AI algorithms. However, non-experts often face challenges in applying these methods, requiring intensive collaboration with data scientists. The resulting knowledge gap necessitates mediators, increasing time and resource demands. Potential solutions include training domain experts in DS, data scientists in the domain, or employing a conversational AI mediator to seamlessly bridge this skills gap.

---

[5] Service Level Objectives (SLOs) provide a quantitative framework for defining the level of service a customer can expect from a provider. SLOs are established by setting specific goals for metrics, referred to as Service Level Indicators (SLIs). For instance, an availability SLO might specify the target value for an availability SLI, measured over a defined period (e.g., four weeks), representing the expected standard of service.
[6] https://liris.cnrs.fr/en/team/bd

Data analysis through AI models must uphold principles of privacy, transparency, and respect for the exchange, sharing, and processing of raw data. AI systems risk undermining these values without careful oversight. In allocating resources for the evaluation of DS queries, several key **technical** aspects must be considered:

- **Computational Power**: Ensure that participating nodes have sufficient resources to effectively prepare data and train and test models.
- **Network Bandwidth**: Efficiently manage data and results transfer among nodes to avoid bottlenecks during task execution.
- **Scalability**: Allocate resources to support the seamless addition of nodes without compromising performance.
- **Latency**: Minimize delays in communication between nodes to maintain efficient model updates and task synchronisation.
- **Storage**: Provide adequate storage for models as well as initial, intermediate, and final datasets throughout the process.

By addressing these aspects, resource allocation can effectively support the responsible and efficient execution of DS queries while safeguarding ethical principles.

Additional **contextual** and **qualitative** constraints can be integrated into the conditions under which a DS query evaluation process is executed to ensure it is conducted responsibly and inclusively. A responsible execution minimises resource and energy consumption, reduces the $CO_2$ footprint, and adheres to region-specific process localisation. It also prioritises data and execution sovereignty, optimises the costs of model training and deployment, and ensures transparency in resource allocation and data management, including storage, transfer, and processing. These attributes collectively guide and influence the execution of DS queries evaluation, ensuring alignment with ethical, sustainable, and regionally appropriate practices.

*Fairness and resources allocation*

The next steps for advancing fairness in resource allocation and analytics processes involve several key actions:

- **Develop comprehensive fairness metrics**: Design robust, multi-dimensional fairness metrics that account for intersectionality, ensuring that overlapping attributes (e.g., age, gender, ethnicity) are adequately reflected in fairness assessments along the input and produced data, content, algorithms and hardware infrastructure.
- **Promote intersectional[7] approaches**: Encourage the integration of intersectional strategies into fairness research and practice, addressing the unique experiences and challenges faced by diverse groups.
- **Integrate fairness into DS systems**: Implement mechanisms for real-time measurement and monitoring of fairness during task execution and resource allocation, enabling dynamic adjustments to maintain equity.
- **Enhance compliance verification**: Create automated tools and frameworks to continuously verify and enforce compliance with fairness metrics across all nodes and federations.
- **Incorporate fairness in decision-making**: Embed fairness principles into algorithms governing process deployment, job scheduling, and resource allocation, ensuring these processes are guided by fairness constraints.

---

[7] Intersectionality is a conceptual framework that examines how different forms of social identities—such as race, gender, class, sexual orientation, ability, and other characteristics—intersect and interact to create unique and overlapping systems of oppression or privilege.

- **Continuous evaluation and adaptation**: Establish ongoing evaluation frameworks to measure the effectiveness of fairness measures and adapt them as new challenges and contexts arise.
- **Foster interdisciplinary collaboration**: Collaborate across disciplines, integrating insights from social sciences, ethics, and technology to develop holistic approaches to fairness in data analytics and resource allocation.

My work aims to provide fair resource allocation for DSs phases through nodes and on models produced and aggregated in DS queries evaluation and thus conform to expected fairness metrics specified by the community producing/consuming data and analytics results.

*Research questions*

Broadly, the research questions we are exploring revolve around the following key areas:

- Defining Fairness Metrics: How can we establish measures that accurately define fairness in resource allocation?
- Measuring Fairness online: What methods can be used to assess fairness during execution, and how this metric can be considered into process deployment and resource allocation strategies once the execution environment (cluster, sets of servers in the fog or edge) is established?
- Ensuring Compliance: How can we verify that individual nodes and the entire execution environment adhere to the established fairness index?
- Respecting Fairness Constraints: What strategies can ensure that deployment decisions, job assignments, and resource allocations align with the fairness constraints the execution environment seeks to uphold?

## Related work

The related work for this project spans several key areas directly aligned with our research objectives. Fair resource allocation in data centres involves distributing resources like CPU, memory, and bandwidth among users to ensure efficiency and fairness. Techniques such as Dominant Resource Fairness (DRF) and its extensions, like DRF-CE, address these challenges by considering multiple resource types and user demands. These methods consider fairness properties such as envy-freeness, Pareto efficiency, and strategy-proofness. Additionally, new mechanisms have been proposed to optimise resource allocation in cloud-edge systems, improving user efficiency by considering bandwidth demand compression and other factors. Additionally, fair data preparation underscores the necessity of using unbiased and representative datasets for training AI models. Responsible AI development, or "sobriety," emphasizes ethical and balanced approaches to training and deploying AI, keeping societal considerations central to innovation. Lastly, feminist and decolonial perspectives in technology offer insights into how gender and power dynamics shape technological development, promoting more inclusive and equitable practices. Collectively, these areas inform our approach to building fair, ethical, and inclusive AI systems.

*Fair data preparation*. Existing techniques strive to ensure fairness in data analysis by addressing bias and promoting the representation of diverse groups characterised by various attribute values within data collections. For example, when analysing a group of people's career development, these techniques ensure that individuals of all ages and genders are fairly represented in the data. In implementing analytics processes, ensuring the fair treatment of data throughout the entire data analysis pipeline, from data collection to processing and analysis is crucial. Fairness in data analytics [9], [10] involves addressing biases, ensuring equitable treatment of all participants, and developing methodologies that promote fairness throughout all data collection, processing, and analysis stages. It is a continuous effort that engages researchers, practitioners, policy makers, and society. Existing work on fairness in data analytics includes approaches that advocate for fairness of data in ways that avoid creating or perpetuating biases [11], [12]. Additionally, these efforts extend to incorporating fairness and justice into data fairness processes through the ethics of data analytics [14], with some works also integrating feminist perspectives [15]. When examining issues of equity, diversity, and inclusion, intersectionality—considering multiple overlapping attributes—is essential for achieving a truly equitable perspective. This recognition motivates the proposal of intersectional strategies that more effectively address these challenges in data analysis.

*Sobriety in AI* model training and deployment emphasises a measured, responsible approach that avoids hype and reckless development. Key aspects of this approach include ethical sobriety, such as thoroughly evaluating potential adverse impacts, incorporating safeguards and human oversight, and prioritising fairness, transparency, and accountability. Rigorous testing and validation play a significant role, ensuring that AI models are thoroughly tested across diverse scenarios before deployment, with ongoing monitoring and a readiness to redesign systems. Thoughtful data practices are essential, including using high-quality, representative datasets, respect for data privacy, and efforts to mitigate biases. Responsible resource use is another key aspect, involving considering environmental impacts, optimising for efficiency, and exploring less computationally intensive approaches.

Emerging approaches use several models with different sizes and performances to address queries with varying requirements of performance. The work in [16] proposes a novel hybrid approach to reduce the computational costs associated with deep neural networks while maintaining performance. The work proposes a cost-effective and quality-aware query routing for LLMs. The authors identify several approaches for reducing ML inference costs, such as model compression techniques (pruning, quantization, and knowledge distillation) and neural architecture search to optimise model architectures.

Hybrid inference paradigms that use two models of different sizes to reduce costs [17], [18]. The core concept is identifying and directing more straightforward queries to a smaller model, thereby reducing inference costs and preserving response quality. Adjusting a threshold based on query complexity makes it possible to balance quality and price within the same inference framework dynamically.

*Feminism and decoloniality in technologies.* Techno-feminism[8] seeks to transform the tech industry by embedding a feminist perspective at every stage of development. This involves increasing female representation in development teams, redesigning processes to meet diverse user needs, and addressing unconscious biases. It promotes inclusive practices and drives innovation, creating technologies that challenge and dismantle existing inequalities, fostering a more equitable digital future.

Decolonisation studies in data and algorithms focus on addressing the systemic biases and inequities embedded in data science and machine learning practices [25]. Key statements emphasise the need to challenge Eurocentric and Western-centric paradigms that dominate data collection, processing, and analysis [9]. These studies advocate for incorporating Indigenous and non-Western epistemologies, respecting data sovereignty, and ensuring diverse and inclusive participation in data-driven projects. Results include the development of frameworks for ethical AI, algorithms that prioritise fairness and accountability, and methodologies that center marginalised voices in decision-making processes.

---

[8] Joy Buolamwini, Techno-féminisme : comment les femmes redéfinissent la technologie pour un avenir plus inclusif, Magazine 1981, https://www.1981-magazine.fr/index.php/2024/08/21/techno-feminisme-comment-les-femmes-redefinissent-la-technologie-pour-un-avenir-plus-inclusif/- Accessed 22nd August 2024
[9] https://aplusalliance.org/about-fair/

Decolonial approaches also highlight the importance of transparency, power-aware system design, and community-driven solutions to foster equitable outcomes in data science and algorithmic practices.

**Background and general objective**

When addressing query evaluation and optimisation, tasks such as aggregation, filtering, traversal, and analytics operations can have varying computing requirements based on the data being processed and the algorithmic complexity involved. The strategies used to allocate computing resources to these tasks significantly impact query performance, particularly in terms of time cost and adherence to service level objectives (SLOs) and economic cost when resources are associated to economic cost models (e.g., the pay as U go model of many public cloud providers). Additionally, qualitative constraints related to task execution, as well as the storage and sharing of input, intermediate, and final results (data and models), must be considered.

It is crucial to evaluate not only the technical execution of DS query tasks but also their qualitative aspects. The results of a DS query should include data, models, and scores, alongside detailed information about the conditions under which these results were obtained and the degree to which qualitative requirements were met. This approach ensures that DS practices move toward being respectful, transparent, and fair, fostering trust and equity in the broader application of these processes. Moreover, curating the conditions under which DS queries are evaluated enables their curation and facilitates the tracking of knowledge associated with their design and implementation in benefit of the collectivity well-being and not seeking economic return of investment.

My work on "just in time architectures", trust and service level agreements (SLA) and service level objectives (SLO)s and data curation will provide a solid background for addressing these challenges (cf. Report of "connaissances antérieures").

> The **primary objective** of this project is to develop fair and responsible protocols for designing conversational DS queries with human-in-the-loop, supported by optimisation algorithms that allocate computing resources while balancing quantitative and qualitative objectives. Fairness and responsibility will be systematically evaluated and embedded across data, DS processes, and system architectures. The curation of evaluation and allocation decision-making processes will ensure traceability, foster knowledge production and reuse, and contribute to advancing "open science" initiatives.

**Towards fair resource allocation with a decolonial and responsible perspective**

Our work focuses on developing a data management approach within DS execution environments to ensure that they are equitable, providing a fair data analysis from a decolonial (i.e. techno-sovereignty) and responsible perspective (see Figure 1).

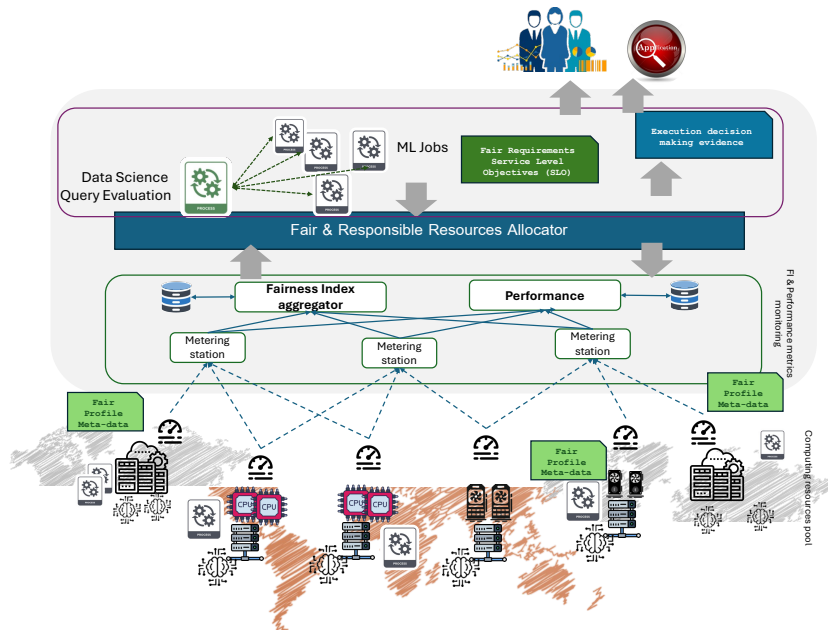

*Figure 1. Fair Resources Allocation for DS Query Evaluation*

Three key challenges must be addressed to integrate this approach into the execution of DS queries: (1) the declaration of target indices that reflect the expectations of the application or user community; (2) the specification of metrics that can be dynamically monitored to assess the extent to which the execution aligns with predefined thresholds; and (3) the reinforcement of the "contracts" established between the DS query execution environment and the users or applications to ensure adherence to these agreed-upon standards.

*Specifying fair indices for resources allocation.*

My vision for defining a fairness index with an intersectional perspective involves combining several key metrics into a weighted formula, allowing for prioritising specific preferences. We propose integrating metrics such as servers' location and provenance, data sovereignty and provenance, model performance, training time, CPU/GPU cores, and calibrating cycles to achieve an optimal global performance within the evaluation phases of a DS query. Each phase includes three essential tasks: dispatching jobs to participating servers, integrating results, and evaluating overall performance. Two additional complex attributes are considered in our approach: (i) training cost, which encompasses the $CO_2$ footprint and economic cost, calculated based on the training time, CPU/GPU cores, and calibrating cycles; and (ii) economic cost, determined by factors such as CPU/GPU cores, storage, and networking availability [24].

$$\begin{aligned}FainessIndex(FI) =& \alpha_1 L_s + \alpha_2 P_s + \alpha_3 DS_p + \alpha_4 MP + \\ & \alpha_5 T_t + \alpha_6 C_{GPU} + \alpha_7 C_{cal} + \\ & \beta_1 TC_{CO2} + \beta_2 EC\end{aligned} \quad (1)$$

- $L_s$ represents servers location,
- $P_s$ represents provenance,
- $DS_p$ denotes data sovereignty and provenance,
- MP indicates model performance,
- $T_t$ is training time,
- $C_{GPU}$ denotes the number of CPU/GPU cores used,
- $C_{cal}$ refers to calibrating cycles,
- $TC_{CO2}$ represents the training cost in terms of $CO_2$ footprint,
- EC denotes the economic cost.
- The coefficients $\alpha_1 ... \alpha_7, \beta_1, \beta_2$ are weights assigned to each metric. They reflect their relative importance within the fairness index. These weights can be adjusted based on the specific priorities or fairness objectives of the federated learning process.

This formula allows for a balanced consideration of various critical factors, ensuring that the DS query evaluation process is fair, efficient, and responsible in terms of both environmental, social and economic impacts. The weights assigned to metrics are determined by the application context. The sum of all coefficients must equal 1.

For each DS query job, it is essential to associate it with a set of preferences related to these attributes. A preference item is expressed as an n-tuple, which pairs the attribute with a threshold value and a preference weight expressed as a percentage. For instance, the following n-tuple might represent a geographical location attribute using four points of geographical coordinates, with a preference of 35%:

```
< location, < northernmost-point,
              southernmost-point,
              easternmost-point,
              westernmost-point  >, 0,35 >
```

For instance, one might specify that servers for a given DS query should be located in Chiapas, Mexico. The location is defined in terms of geographical coordinates, outlining a specific geographic region, while provenance involves selecting servers from a predefined set of acceptable owners. Additionally, data sovereignty should be ensured through a certification process for the servers participating in the execution of a DS query. The attributes training time and CPU/GPU cores can be used to estimate the training cost, particularly regarding the $CO_2$ footprint associated with model training. This involves calculating the Carbon Intensity of Electricity (in grams of $CO_2$ per kWh), which varies by region and is influenced by the local energy mix (e.g., coal, gas, renewables).

*Measuring and monitoring fairness metrics*

The principle of a DS architecture is that participating servers operate independently, functioning as almost black boxes concerning the specifics of the data used for training and testing models. This approach ensures that the results do not inadvertently reveal any information about the underlying data. To uphold this level of privacy and security, it is crucial to develop strategies to collect and estimate fairness metrics associated with these servers without compromising data confidentiality. Additionally, other relevant metrics can be provided through metadata associated with the servers that opt to participate in the DS architecture, further enhancing the transparency and accountability of the process. We build upon our previous work [2], [3], focused on estimating trust metrics—such as data freshness and update frequency—from black box data services. In these environments, where direct access to the underlying data is restricted, we developed methods to infer these metrics based on observable outputs and interaction patterns. This approach allows us to assess the reliability and timeliness of the data provided by these services, even when the internal processes and data management practices are not transparent. By extending these techniques to the context of DS queries evaluation, we aim to estimate similar trust metrics for participating servers, ensuring that they contribute valuable and up-to-date information to the learning process without compromising the privacy and autonomy of the individual datasets. To ensure analytics sovereignty in using AI models for addressing research questions specific to communities or issues in the Global South, it is essential to collect detailed data about the conditions under which the analytics process is conducted. This data should encompass the number of cycles required for convergence to a global result, the parameters for calibrating the international model, the selection and number of participating servers, and their adherence to fairness metrics. Additionally, it is essential to gather information on the worldwide model itself, including its performance metrics, associated findings, and the environmental and economic costs incurred. This comprehensive approach helps maintain control over the analytics process, ensuring that it remains aligned with the interests and needs of the communities it serves.

*Responsible resources allocation in target computing architectures*

Existing DS environments are typically "one-size-fits-all" cloud systems. They address the analytics and data management divide by offering integrated backends that efficiently execute analytics pipelines while allocating the necessary infrastructure resources, such as CPUs, FPGAs, GPUs, and TPUs, alongside platform services like Spark and TensorFlow. These environments are designed to support the execution of DS tasks that demand significant storage and computational resources, with pipelines that often transition from in-house executions to cloud deployments. This shift necessitates elastic architectures that dynamically scale resources to meet varying demands. Disaggregated data centre solutions offer a promising approach to meeting these needs.

Disaggregated data centres are emerging as promising alternatives to fulfil the requirements of these applications. They promote composable infrastructures that can be customised to dynamically provide resources according to the workloads submitted by applications and fulfil their Service Level Objectives (SLOs). Our previous work has focused on addressing the challenges of coupling disaggregated data centre architectures with the execution of DS workflows, aiming to optimize resource allocation and efficiency [4].

We aim to refine our resource dispatching algorithm SLO's dynamically meet SLOs such as performance and energy consumption while adhering to fairness constraints. By incorporating a fairness specification based on the previously mentioned metrics and optimizing for the Fairness Index (FI), the dispatching algorithm will allocate computing resources and delegate the execution of analytics tasks in alignment with the SLOs and FI. As the algorithm monitors the execution status of jobs assigned to various servers and the overall progress of the DS queries evaluation, it can dynamically adjust resource allocation. For example, the algorithm might allocate GPU resources to train a model calibrated to prioritize the FI, which could involve lowering performance expectations to reduce energy consumption. Subsequently, it could switch to allocating CPU resources for the testing phases or model integration tasks at the global level within the DS architecture.

This dynamic approach ensures that resource allocation is continuously optimised to balance fairness, performance, and energy efficiency. Since the optimisation objectives of both SLO and FI can potentially not be achieved, a negotiation strategy applied and develop a best-effort optimisation approach for implementing the resources allocation strategy. In this sense, we will elaborate on our current work regarding trust in distributed systems [5].

*Open science*

We have extended the concept of curation to DS queries, proposing that every DS query constitutes a data-driven experiment addressing a research question from another discipline. Our approach focuses on curating both quantitative and qualitative experimental methodologies. This involves capturing static aspects—such as data, scientists, and analytical tools—and dynamic aspects, including the decision-making processes that guide an experiment through its phases, from validation and refinement to dissemination of results. The key contribution is a set of versioning operators, guided by consensus, to manage and organise the content generated during data-driven experimental processes, ensuring traceability and reproducibility.

## Organization and expected results

Existing work on responsible resource allocation strategies for DS queries reveals several limitations and open issues. While significant progress has been made in fairness-aware DS tasks, fair resource allocation, and equitable data preparation, these approaches often fail to address the deeper complexities of power dynamics, resource inequities, and biases embedded in DS systems. Current methods predominantly focus on technical aspects, such as workload distribution but frequently neglect broader socio-economic implications, including the reinforcement of existing power structures and the risk of epistemic violence.

Although there is growing recognition of the importance of incorporating principles such as sobriety in model training and feminist perspectives, comprehensive frameworks that holistically integrate fairness, equity, and inclusivity across all stages of DS workflows remain scarce. These gaps underline the urgency for interdisciplinary approaches and the development of strategies that prioritise the voices and needs of decolonialised and marginalised communities. Such frameworks must ensure that the benefits of AI technologies are equitably distributed.

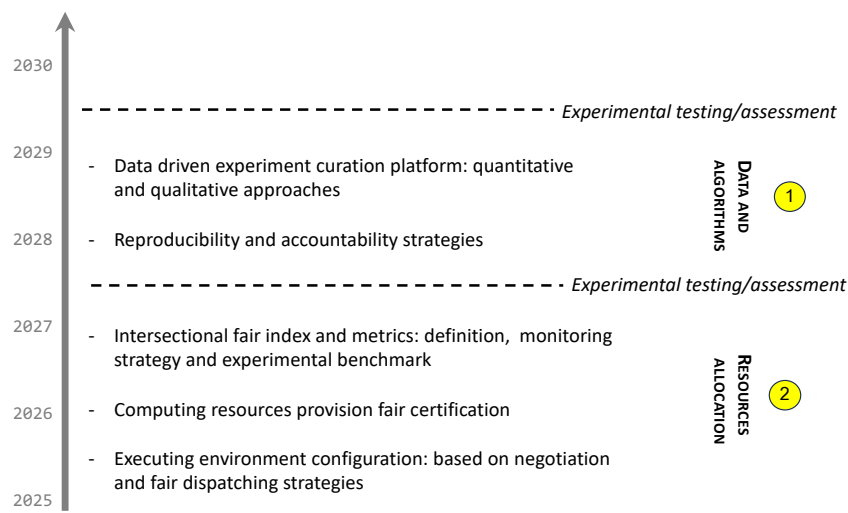

Figure 2. Organisation timeline

Implementing these principles requires meaningful community engagement, interdisciplinary collaboration, and a willingness to challenge dominant paradigms in AI and ML research. The aim is to create DS practices that empower communities rather than exploit them for value extraction. Grounded in a decolonial perspective, our initial studies emphasize that fair resource allocation in DS queries evaluation necessitates a critical examination of power dynamics and a focus on addressing the needs of marginalised groups. Developing practical solutions will require addressing key considerations, including (see Figure 2):

1. Data and Algorithms
   - **Representation and Participation**: Ensure diverse and equitable participation from communities historically excluded from AI/ML development, particularly those in the Global South. This involves actively collaborating with and empowering local organisations, researchers, and stakeholders. My work with the groups of the Feminist A+I network of the A+I alliance will provide a credible context for connecting with local organisations.
   - **Data Sovereignty**: Uphold the principles of Indigenous data sovereignty, ensuring communities maintain ownership, control, and decision-making authority over their data rather than having it extracted for centralised or external purposes.
   - **Pluriversal Approach**: Embrace diverse ways of knowing and culturally grounded approaches to machine learning, moving away from a one-size-fits-all model and allowing for solutions tailored to the needs of specific communities.
   - **Local Context**: Design DS queries evaluation systems that are adaptable to local contexts, languages, and knowledge systems. These systems should respect and incorporate local frameworks rather than imposing Western or Global North paradigms.

2. Computing resources allocation in different target architectures
   - **Transparency**: Ensure that decision-making processes related to resource allocation are fully transparent and accountable to all participants, fostering trust and inclusivity within the system.
   - **Decolonial Metrics**: Develop alternative metrics for evaluating fairness and performance that go beyond traditional Western standards of accuracy and efficiency. These metrics should incorporate cultural relevance, sustainability, and social impact, aligning with diverse community values and priorities.

To create a truly equitable DS, it is vital to ensure that its benefits and fairness are distributed equitably across all participating communities, rather than being concentrated within Big Tech companies or research institutions. Achieving this requires substantial investment in capacity building, especially in underserved regions, where local machine learning and AI capabilities must be strengthened instead of exploiting these areas solely for data and computational resources. Additionally, the design of DS architectures must be power-aware, critically assessing whether these systems perpetuate existing power imbalances or actively work to redistribute power more fairly.

To support this shift, it is imperative to centre Indigenous and non-Western ethical frameworks in the design and governance of these systems. This approach ensures that diverse cultural perspectives and values are respected and meaningfully incorporated. By embedding these principles, DS can become a more inclusive, fair, and responsive field, addressing the needs and priorities of all stakeholders while fostering equity and justice in its practices and outcomes.

**Project implementation strategy**

To implement the proposed research project effectively, the following strategies will be pursued:

1. **Securing Funding**: Actively seek funding from both national and international research grants, including CNRS programs, Horizon Europe, and interdisciplinary initiatives focusing on AI ethics, sustainability, and decolonial methodologies. Tailor proposals to align with the specific criteria of each funding body, emphasising the project's innovative and transdisciplinary aspects.

2. **Building Partnerships**: Collaborate with academic institutions, research centers, and NGOs with shared interests in responsible and inclusive data science. Engage with industrial partners, especially those committed to ethical AI and sustainable practices, to co-develop tools and methodologies. Utilise networks such as the Feminist A+I alliance and other interdisciplinary collaborations for broader reach and synergy.

3. **Attracting PhD and Postdoc Fellows**: Leverage the project's alignment with cutting-edge trends like fairness-aware AI and decolonial data science to attract talented early-career researchers. Promote opportunities through academic networks, conferences, and specialised forums. Provide mentorship, training, and career development support, emphasising the opportunity to contribute to socially impactful research.

4. **Interdisciplinary Engagement**: Foster collaboration between computer scientists, ethicists, and social scientists to address the project's multifaceted challenges. Encourage participation from diverse backgrounds, prioritising underrepresented groups to enrich the research perspective.

5. **Global Outreach**: Develop partnerships with institutions and researchers in the Global South to ensure diverse representation and perspectives in the research process. Strengthen regional AI capabilities by hosting workshops, providing training, and co-developing locally relevant solutions.

6. **Open Science Initiatives**: Commit to transparency and reproducibility by sharing data, algorithms, and results in open-access platforms. This aligns with the project's focus on fairness and inclusivity and contributes to building trust within the broader scientific community.

These strategies aim to create a robust foundation for the project's successful execution while fostering a collaborative, inclusive, and impactful research environment.

A promotion to a CNRS Research Director position would provide the leadership, resources, and visibility needed to drive the proposed project's ambitious goals. This role can enhance access to funding, advanced infrastructure, and high-profile collaborations, particularly with international and Global South partners, amplifying the project's reach. It can strengthen the ability to attract top-tier PhD students and postdocs, fostering a vibrant research environment. By championing inclusive, fair, and decolonial methodologies within CNRS and global networks, the position could enable impactful, interdisciplinary research that advances innovation, equity, and societal contributions.